\documentclass[12pt]{iopart}
\usepackage{iopams}
\usepackage{setstack}

\usepackage[T1]{fontenc} 
\usepackage[utf8]{inputenc}	
\usepackage[final,babel]{microtype} 

\usepackage[main=english]{babel}	
\usepackage{url} 
\usepackage[usenames,dvipsnames,svgnames,table]{xcolor} 

\usepackage{bm} 
\usepackage{braket}
\usepackage{isotope} 
\usepackage{siunitx} 
\DeclareSIUnit\gauss{G} 
\newcommand{\eff}{ef\hspace{-0.5mm}f}

\usepackage{graphicx} 
\usepackage{epstopdf} 
\usepackage{wrapfig} 
\usepackage{fancybox} 

\RequirePackage{hyperref} 
\usepackage{hyperref} 
\hypersetup{
  linkcolor=blue,
  filecolor=blue,
  urlcolor=blue,
  citecolor=blue,
  bookmarksnumbered=true,
  bookmarksopen=true,
  bookmarksopenlevel=1,
  pdftitle={Depletion Imaging of Rydberg atoms in cold atomic gases},
  pdfstartview={FitH} 
}

\begin{document}
\title[Depletion Imaging of Rydberg atoms in cold atomic gases]{Depletion Imaging of Rydberg atoms in cold atomic gases}

\author{M.~Ferreira-Cao$^1$, V.~Gavryusev$^1$, T.~Franz$^1$, R.~Ferracini Alves$^1$, A.~Signoles$^1$, G.~Z{\"u}rn$^1$ and M.~Weidem\"{u}ller$^{1,2}$,}
\address{$^1$Physikalisches Institut, Universit\"{a}t Heidelberg, Im Neuenheimer Feld 226, 69120 Heidelberg, Germany.}
\address{$^2$Hefei National Laboratory for Physical Sciences at the Microscale and Department of Modern Physics, and CAS Center for Excellence and Synergetic Innovation Center in Quantum Information and Quantum Physics, University of Science and Technology of China, Hefei, Anhui 230026, China. }

\date{\today}

\ead{\mailto{ferreira@physi.uni-heidelberg.de}, \mailto{weidemueller@uni-heidelberg.de}}

\begin{abstract}
We present a depletion imaging technique to map out the spatial and temporal dependency of the density distribution of an ultracold gas of Rydberg atoms. Locally resolved absorption depletion, observed through differential ground state absorption imaging of a~\isotope[87]{Rb} cloud in presence and absence of pre-excited Rydberg atoms, reveals their projected two-dimensional distribution. By employing a closed two-level optical transition uncoupled from the Rydberg state, the highly excited atoms are preserved during imaging. We measure the excitation dynamics of the $\ket{48S}$ state of $^{87}\text{Rb}$, observing a saturation of the two-dimensional Rydberg density. Such outcome can be explained by the Rydberg blockade effect which prevents resonant excitation of close-by Rydberg atoms due to strong dipolar interactions. By combining the superatom description, where atoms within a blockade radius are represented as collective excitations, with a Monte Carlo sampling, we can quantitatively model the observed excitation dynamics and infer the full three-dimensional distribution of Rydberg atoms, that can serve as a starting point for quantum simulation of many-body dynamics involving Rydberg spin systems.
\end{abstract}

\vspace{1pc}
\vspace{1pc}


\section{Introduction}
\label{sec:intro}

The excitation dynamics of Rydberg atoms in dense atomic clouds has been found to present strong deviations from the single-particle regime due to the effect of long-range Rydberg-Rydberg interactions, which strongly suppress the excitation probability in the blockaded regime~\cite{Tong2004,Singer2004,Gallagher2008,Comparat2010} and lead to collective dynamics~\cite{Gaeetan2009,Urban2009,Barredo2014,Helmrich2018}.

In one- and two-dimensional discrete geometries, spatially resolved Rydberg excitation dynamics has been observed by combining deterministic preparation and fluorescence detection of atoms in optical lattices or tweezer arrays~\cite{Barredo2014,Schauss2015,Levine2018}. This allowed to investigate the many-body dynamics in quantum Ising spin systems~\cite{Labuhn2016,Bernien2017,GuardadoSanchez2018}.
In continuous three-dimensional systems, global Rydberg properties ~\cite{Reetz-Lamour2008,Loew2009,Dudin2012a,Helmrich2018} and two-dimensional integrated spatial profiles have been used to study the excitation dynamics under the influence of Rydberg-Rydberg interactions. The latter has been explored by means of tomographic techniques based on field-ionization of the Rydberg states and ion counting on an MCP detector~\cite{Valado2013} and by locally resolving the autoionization of a Rydberg cloud~\cite{Lochead2013} or of an ultracold plasma of Rydberg atoms~\cite{McQuillen2013}. Another approach, called Interaction Enhanced Imaging~\cite{Guenter2012,Guenter2013,Gavryusev2016}, reveals the presence of down to a few Rydberg excitations by using the surrounding ground state atoms under electromagnetically induced transparency coupling as a contrast medium. In the non-interacting regime, it has been possible to reconstruct the three-dimensional Rydberg distribution and even the density matrix from joint measurements of the local optical spectrum and of the excited atom number~\cite{Gavryusev2016a}.

Here, we report on a complementary scheme to determine the integrated two-dimensional density profile of a Rydberg atom distribution by a fully optical technique, to which we refer as depletion imaging. By leveraging standard resonant absorption imaging~\cite{Ketterle1999} on a closed two-level optical transition that is uncoupled from any Rydberg state, we leave the Rydberg population unperturbed and we detect their presence by comparing the cloud absorption in presence and absence of pre-excited Rydberg atoms. Alternative dispersive imaging techniques based on an off-resonant probing of the atomic ensemble, like dark-ground, phase-contrast or Faraday imaging~\cite{VestergaardHau1998,Bradley1997,Gajdacz2013}, could be used for non-destructive measurements of the cloud and stable atom number preparation, with the disadvantage of providing a significantly lower SNR. We present the experimental method and explore its advantages and limitations. Using this method, we perform a spatially resolved measurement of the local Rydberg excitation dynamics, which reveals the saturation of the Rydberg fraction to a value significantly below the expectations for single particle excitation. This constitutes a signature of the Rydberg blockade effect~\cite{Comparat2010}. To quantitatively model the dynamics, we use a Monte Carlo superatom model, where atoms within a blockade radius are described as collective excitations, finding good agreement with the experimental observations for $3.6$ atoms per blockaded volume. By applying this model we reproduce discrete three-dimensional distributions of Rydberg atoms that can serve as an input to study many-body Rydberg spin dynamics~\cite{Schachenmayer2015a,Signoles2019}.

\section{Experimental observation of Rydberg ensembles by depletion imaging}
\label{sec:setup}


\subsection{Preparation of ultracold Rydberg atomic ensembles}
\label{sec:cloudpreparation}

Experimentally, we create an ultracold ensemble of~\isotope[87]{Rb} atoms in a crossed optical dipole trap~\cite{Hofmann2014}, obtaining a three-dimensional cigar shaped cloud of atoms in their electronic ground state $\ket{5S_{1/2}}$ at a temperature of $\SI{40}{\micro\kelvin}$. The atoms are first optically pumped into the Zeeman manifold $F=1$, and then prepared into the hyperfine sublevel $\ket{g} = \ket{5S_{1/2},F=2,m_F=2}$ by a microwave rapid adiabatic passage. To lift the degeneracy between the sublevels, a $6.1(1)$~G magnetic field is applied along the $\hat{z}$ direction, thereby defining the quantization axis. 

\begin{figure}[!!ht]
\centering
\includegraphics[width = 0.8\linewidth]{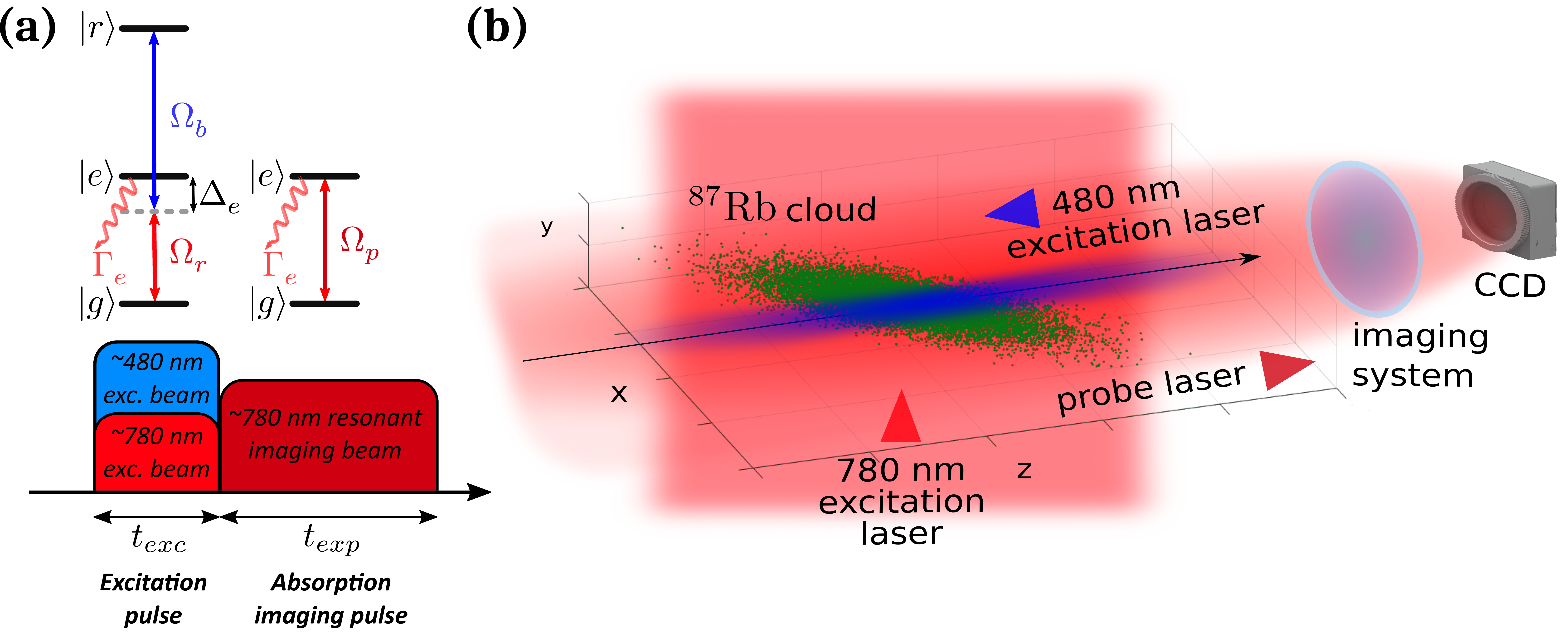}
\caption[Experimental observation of Rydberg atoms by depletion imaging.]{Experimental observation of Rydberg atoms by depletion imaging. 
\textbf{(a)} Excitation scheme and experimental sequence. The two-photon excitation pulse is applied during a variable time $t_{exc}$, then resonant absorption imaging is performed with fixed exposure time $t_{exp}=\SI{5}{\micro\second}$.
\textbf{(b)} Geometry of the experimental setup. The cloud of~\isotope[87]{Rb} atoms in their electronic ground state is prepared in an ellipsoid shape by an optical dipole trap. A vertical $\SI{780}{\nano\meter}$ laser beam is used for Rydberg excitation, in combination with an horizontal $\SI{480}{\nano\meter}$ laser beam, tilted by $\SI{45}{\degree}$ with respect to the dipole trap axis. To measure absorption, a counter-propagating probe laser beam uniformly illuminates the cloud and the transmitted photons are collected onto a CCD camera through an imaging system. 
}
\label{fig:setup}
\end{figure}

An ultracold Rydberg gas is produced by releasing the ground state atoms from the dipole trap and performing an optical two-photon excitation from the ground state to the desired Rydberg state. Here, we address a zero angular momentum state of principle quantum number $n=48$ by combining a $\SI{780}{\nano\meter}$ and a $\SI{480}{\nano\meter}$ photon (see Fig.~\ref{fig:setup}(a)). The red laser beam drives the transition between the $\ket{g}$ state and the intermediate short-lived state $\ket{e}=\ket{5P_{3/2},F=3,m_F=3}$, while presenting a homogeneous Rabi frequency $\Omega_{r}$ over the cloud and a detuning of $\Delta_e/2\pi=\SI{-97.0(4)}{\mega\hertz}$. The blue laser beam drives the transition from $\ket{e}$ to the Rydberg state $\ket{48S_{1/2}}$, while compensating for the detuning of the red photon, and has a spatially-dependent Rabi frequency $\Omega_b(x,y)$. Due to the Zeeman effect, the two fine structure sublevels $m_j=\pm 1/2$ of the Rydberg state are split by $\simeq\SI{17.1(3)}{\mega\hertz}$. We selectively address the fine state $\ket{r}=\ket{48S_{1/2}, m_j=1/2}$ by setting the two-photon frequency to be on resonance with this state.

The polarizations of the laser beams are chosen to address only the three atomic levels $\ket{g}$, $\ket{e}$ and $\ket{r}$. The $\SI{780}{\nano\meter}$ excitation beam propagates along the vertical direction with a linear polarization perpendicular to the quantization axis (see Fig.~\ref{fig:setup}(b)), resulting in a combination of $\sigma^+$ and $\sigma^-$ polarizations, while the blue laser propagates along the quantization axis with a pure $\sigma^-$ polarization. Hence, only the $\sigma^+$ component of the red beam contributes to the two-photon excitation process according to the selection rules. 
We avoid populating the intermediate state $\ket{e}$ by selecting the detuning $\Delta_e$ to be much larger than the one-photon Rabi frequencies $\Omega_r,\,\Omega_b$ and than the intermediate state decay rate $\Gamma_e/2\pi=\SI{6.07}{\mega\hertz}$. Consequently, this scheme may be considered as an effective two level transition between $\ket{g}$ and $\ket{r}$, with spatially-dependent effective Rabi frequency $\Omega_{\eff}(x,y)=\Omega_r\Omega_b(x,y)/2\Delta_e$~\cite{Reiter2012}. Both laser beams are switched on and off simultaneously and the number of prepared Rydberg atoms is controlled by varying the excitation time $t_{exc}$. 


\subsection{Depletion imaging of Rydberg atoms}
\label{sec:depletionimaging}

In order to characterize the Rydberg atom distribution produced by the excitation process, we realize an optical imaging scheme based on absorption imaging~\cite{Ketterle1999}, that we call depletion imaging. Absorption profiles of atoms in the ground state $\ket{g}$ are obtained by shining on the atomic cloud a $\sigma^+$-polarized probe laser resonant with the $\ket{g} \leftrightarrow \ket{e}$ transition (Fig.~\ref{fig:setup}(b)). The Gaussian probe beam is collimated with a waist of $\SI{1.5}{\milli\meter}$ (at $1/e^2$), much larger than the typical cloud size, allowing to assume that the atoms are illuminated with a uniform Rabi frequency $\Omega_p/2\pi=\SI{1.57(3)}{\mega\hertz}$, calibrated using the saturated absorption imaging method~\cite{Reinaudi2007}. The transmitted light intensity $I$ is collected onto a CCD camera via a nearly diffraction-limited imaging system, with a resolution of $\SI{4.8}{\micro\meter}$ (Rayleigh criterion). The imaging system magnification factor of $7.7$ leads to an effective single pixel area of $a_{px}=\SI{4.28}{\square\micro\meter}$ in the object plane. The ratio of the transmitted intensity in presence ($I_{g}$) and absence ($I_{ref}$) of atoms provides the cloud absorption profile $A= 1-I_{g}/I_{ref}$, from which the two-dimensional density distribution is obtained by
\begin{equation}
    n^\mathrm{2D}(x,y) = -\sigma_0^{-1} (1+s_0) \ln [1-A(x,y)],
    \label{eq:density2D}
\end{equation}
that corresponds to the three-dimensional atom density $n$ integrated over the imaging direction ($\hat{z}$) $n^\mathrm{2D}(x,y) = \int n(x,y,z)dz$. Here, $\sigma_0$ is the absorption cross section of the imaging transition and $s_0=2\Omega_p^2/\Gamma_e^2$ accounts for intensity saturation effects. The atom number in each pixel is found by multiplying the local density by the pixel area.

\begin{figure}[!!ht]
\centering
\includegraphics[width=0.8\linewidth,keepaspectratio=true]{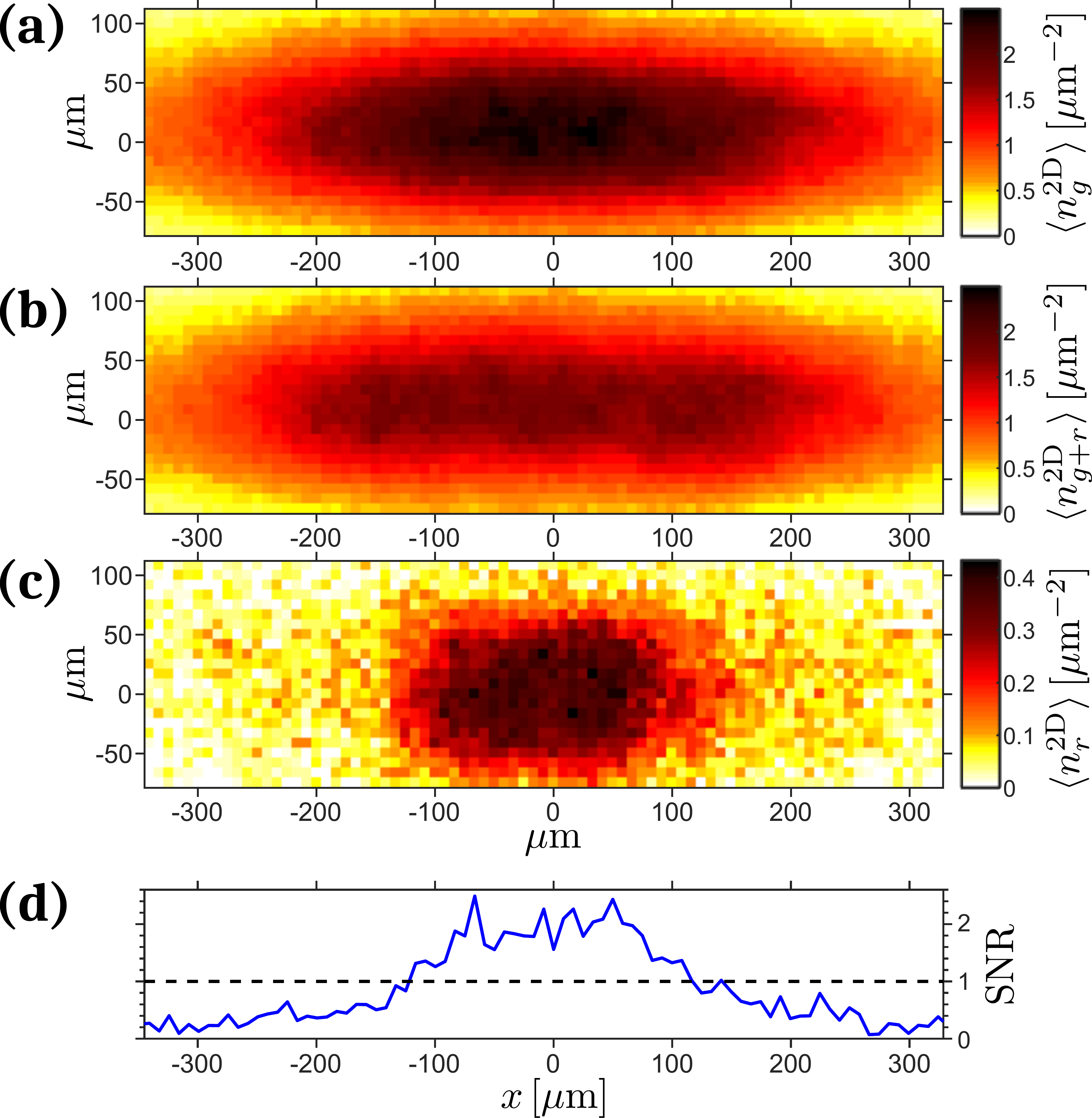}
\caption[Typical two-dimensional density distributions observed during depletion imaging experiments] {Typical two-dimensional density distributions observed during depletion imaging experiments, averaged over 50 realizations and with $16\,a_{px}$ binning. \textbf{(a)} Ground state density without atoms excited to the Rydberg state, obtained from the optical density $\mathrm{OD}=1-A$ with a peak $\mathrm{OD}=0.7$. \textbf{(b)} Ground state density depleted by atoms pre-excited to the Rydberg state for $t_{exc}=\SI{3}{\micro\second}$. \textbf{(c)} Rydberg density for  $t_{exc}=\SI{3}{\micro\second}$ (peak $\mathrm{OD}=0.2$) obtained by depletion imaging through Eq.~\ref{eq:density2D} from the difference in cloud absorption between the realizations presented in (a) and (b). \textbf{(d)} Signal-to-noise ratio within the Rydberg density distribution, along a cut through $y=0$, determined from the fluctuations of the individual realizations.}
\label{fig:2Ddensity}
\end{figure}

Absorption images performed without Rydberg pre-excitation reveal the two-dimensional density distribution $n^\mathrm{2D}_g$ of the ground state atoms after their release from the dipole trap (Fig.~\ref{fig:2Ddensity}(a)). We experimentally measure a Gaussian-shaped distribution with widths $\sigma_x=\SI{233}{\micro\meter}$ and $\sigma_y=\SI{57}{\micro\meter}$ (at $1/e^2$), with an average total atom number of $\langle N_g\rangle \approx 1.27(5)\cdot10^{5}$. Considering that the imaging is performed with a $45^{\circ}$ incidence angle with respect to the dipole trap axis (see Fig.~\ref{fig:setup}(b)), the actual width of the atomic cloud along the longitudinal direction is increased by a factor of $\sqrt{2}$. Furthermore, since the atomic cloud is expected to present a cylindrical symmetry around the axis of the dipole trap laser beam, the three-dimensional ground state density distribution $n_g$ can be deduced from the images.

When taken after the Rydberg excitation step, absorption images reveal the two-dimensional densities $n^\mathrm{2D}_{g+r}(x,y)$ of atoms that remain in the ground state $\ket{g}$. As shown in Figure~\ref{fig:2Ddensity}(b) for an excitation duration of $t_{exc}=\SI{3}{\micro\second}$, a region with a smaller density is clearly visible around the cloud center. We attribute such difference to the reduction of ground state absorbers due to their one-to-one conversion into Rydberg atoms during the excitation step. This relation allows to measure the two-dimensional spatial distribution of the Rydberg ensemble $n^\mathrm{2D}_r \equiv n^\mathrm{2D}_g - n^\mathrm{2D}_{g+r}$, which we determine from a subset of the intensity profiles acquired in the experimental realizations with ($I_{g+r}$) and without ($I_{g}$) Rydberg state pre-excitation, hence realizing depletion imaging. By using equation~\ref{eq:density2D} with the absorption depleted by the reduction of ground state absorbers $A_\mathrm{dep}= 1-I_{g}/I_{g+r}$, we obtain the two-dimensional Rydberg density distribution $n^\mathrm{2D}_r$. Since our excitation scheme behaves as an effective two-level system, the ratio between the Rydberg and ground state densities allows to extract the local Rydberg fraction $\rho_{r}^\mathrm{2D}(x,y)=n^\mathrm{2D}_{r}(x,y)/n^\mathrm{2D}_g(x,y)$.

Figure~\ref{fig:2Ddensity}(c) shows a typical two-dimensional Rydberg density distribution observed through depletion imaging for $t_{exc}=\SI{3}{\micro\second}$. This profile is not purely Gaussian, with a root-mean-square width along the $x$ direction of $\sigma_{x}^{r}=\SI{156}{\micro\meter}$, and it strongly deviates from the ground state cloud shape due to the geometry of the excitation laser beams. Although the red beam illuminates uniformly the atoms, the blue beam is focused on the cloud center, effectively defining the Rydberg excitation volume (see Fig.~\ref{fig:setup}(b)). By integrating the two-dimensional Rydberg density, we measure a total number of Rydberg atoms of $\langle N_r\rangle = 1.68(3)\cdot10^4$, corresponding to an overall excited fraction of $13\%$.


\subsection{Advantages and limitations of depletion imaging}
\label{sec:limitations}

Depletion imaging allows to obtain spatially resolved information on the Rydberg distribution. Since the Rydberg state is completely uncoupled from the $\ket{g}\leftrightarrow\ket{e}$ transition, the resonant absorption imaging detection step realizes a projective measurement that leaves the highly excited atom population intact, despite perturbing the atomic coherences. In principle, such probing could be repeated multiple times to observe the dynamics of the Rydberg ensemble as long as cloud expansion due to ground state heating does not cause a severe loss of contrast.

Depletion imaging features the same limitations as absorption imaging in terms of noise contributions, as both rely on camera based detection of the probe light transmitted by the cloud. In particular, the detection process is affected by the intrinsic photon shot noise, originating from the Poissonian fluctuations of the incident laser light, and from two technical noise sources: the electronic noise introduced by the charge readout on the sensor and the fluctuations of the mean laser intensity between images within the same set. The optimal detection regime for each image is when the photon shot noise contribution dominates over the other sources, allowing for a square root scaling of the signal-to-noise ratio (SNR) with an increasing number of detected photons. By computing the two-dimensional Rydberg density distribution directly from the intensity profiles (using equation~\ref{eq:density2D} with $A_\mathrm{dep}= 1-I_{g}/I_{g+r}$) rather than from the difference between the ground state density distributions without and with Rydberg pre-excitation, depletion imaging avoids introducing the noise contribution of the reference intensity patterns $I_{ref}$, allowing to keep the SNR at a level comparable to standard absorption imaging.

To efficiently detect the Rydberg atoms in the state and spatial configuration they are prepared, the imaging duration must remain short compared to their lifetime $\tau_{48S}=\SI{56}{\micro\second}$, accounting for the spontaneous and black body radiation induced decays. Therefore, we use a short exposure time of $t_{exp}=\SI{5}{\micro\second}$. To ensure that we operate in the photon shot noise limited regime, while not saturating the transition, we choose the probe Rabi frequency to be $\Omega_p/2\pi=\SI{1.57(3)}{\mega\hertz}$. Under these conditions, each atom absorbs on average $\sim 10$ photons, an amount comparable to the noise level, rendering the probability to detect a single ground state atom very small. When several atoms contribute to the absorption observed in one pixel, a sufficient SNR can be obtained in a single acquisition to detect the density distribution of ground state atoms, but not of the smaller Rydberg ensembles produced in the experiment. Therefore, before mapping out the two-dimensional Rydberg density distribution, we average over $N=50$ experimental realizations to reduce the noise impact by a factor of $\sim 7$ and, towards the same goal, we perform a software binning of $4 \times 4 = 16$ pixels, corresponding to an increased effective single bin detection area of $16\,a_{px} = \SI{68.5}{\micro\meter^2}$, albeit at the price of a decreased spatial resolution. Additionally, averaging over many experimental realizations introduces an additional noise source represented by random variations in the cloud absorption level due to fluctuations in both the initial number of ground state atoms (estimated to be within $4\%$) and the number of excited Rydberg atoms. Nevertheless, those two measures allow to obtain an SNR above 1 over most of the Rydberg cloud, as shown in Fig.~\ref{fig:2Ddensity}(d) for a horizontal slice through $y=0$. The boundary with $\textrm{SNR} = 1$ defines the sensitivity $\mathcal{S}$ of our technique, which here is $\sim 15$ Rydberg atoms localized in a single bin. Such SNR enables the spatially resolved study of the Rydberg excitation dynamics in the atomic cloud presented in the next section. Eventually, the SNR could be increased by optimizing the key imaging parameters $\Omega_p$ and $t_{exp}$, while avoiding excessive saturation of the optical transition. With this gain in SNR one could also reduce the size of the binning of the camera pixels and thus improve the spatial resolution. 

\section{Spatial mapping of Rydberg excitation dynamics}
\label{sec:excdynmap}

We apply the depletion imaging technique to characterize the local Rydberg excitation dynamics of the effective two-level $\ket{g} \leftrightarrow \ket{r}$ transition. By performing measurements for increasing excitation times ranging from $t_{exc}=\SI{0.1}{\micro\second}$ up to $\SI{3}{\micro\second}$, we obtain spatially-resolved maps of their two-dimensional distribution and track its time evolution. 
The sum of the duration of the excitation and imaging pulses is much shorter than the Rydberg state lifetime ($\tau_{48S}=\SI{56}{\micro\second}$), so the decay towards the ground state and the redistribution to neighbouring states can be disregarded. Since the off-resonant $\SI{780}{\nano\meter}$ excitation is active during a short time, additional dissipation due to heating of the atomic cloud during $t_{exc}$ can also be neglected. 

\begin{figure}[!!ht]
\centering
\includegraphics[width=1\linewidth,keepaspectratio=true]{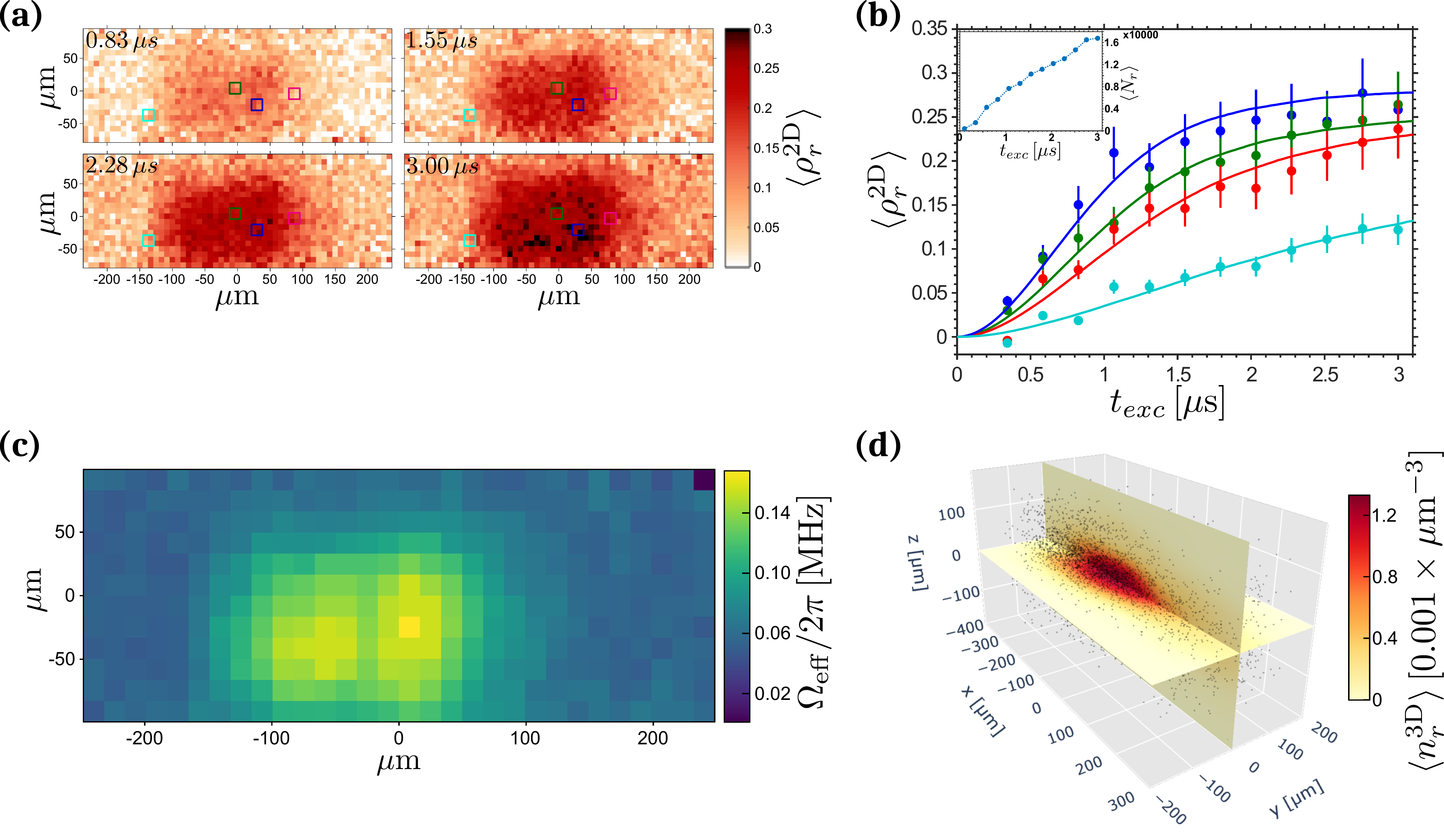}
\caption[Spatially resolved Rydberg excitation dynamics.]{Spatially resolved Rydberg excitation dynamics, averaged over 50 repetitions with $16\,a_{px}$ binning. \textbf{(a)} Two-dimensional Rydberg fraction distribution $\rho^\mathrm{2D}_r(x,y)$, integrated over $\hat{z}$, for increasing excitation times of $0.83$, $1.55$, $2.28$ and $\SI{3.00}{\micro\second}$. \textbf{(b)} Local excitation dynamics at different cloud positions, marked with hollow squares on the $\rho^\mathrm{2D}_r(x,y)$ distribution in \textbf{(a)}. The solid lines represent the simulated dynamics with the master equation Monte Carlo model, using a dephasing of $\gamma/2\pi = \SI{0.25}{\mega\hertz}$ and effective Rabi frequencies $\Omega_{\text{eff}}/2\pi = \SI{0.17}{\mega\hertz}$ (blue line), \SI{0.14}{\mega\hertz} (green line), \SI{0.12}{\mega\hertz} (red line) and \SI{0.07}{\mega\hertz} (turquoise line). Inset: evolution of the global mean Rydberg atom number with $t_{exc}$. \textbf{(c)} Effective Rabi frequency distribution $\Omega_{\text{eff}}(x,y)$, obtained by locally fitting the excitation dynamics with the Monte Carlo model within each $64\,a_{px}$ bin. \textbf{(d)} Reconstructed 3D distribution of the Rydberg atom cloud by the Monte Carlo simulations for $t_{exc}=\SI{3}{\micro\second}$. The dots represent a discrete Rydberg atom arrangement obtained in a single run, whereas the continuous distribution corresponds to the Rydberg density averaged over 50 realizations.}
\label{fig:interaction_effects}
\end{figure}

The experimentally observed Rydberg fraction distribution $\rho_{r}^\mathrm{2D}(x,y)$ is shown in Fig.~\ref{fig:interaction_effects}(a) at different excitation times, where each image bin reveals the dynamics for the local Rabi frequency $\Omega_{\text{eff}}(x,y)$. Further insight into the excitation dynamics can be obtained by exploring the local temporal evolution in different areas of the sample, as depicted in Figure~\ref{fig:interaction_effects}(b), where several regimes are represented: whereas the Rydberg population rapidly increases in the center of the cloud (blue), reaching saturation at $\rho_{r}^\mathrm{2D}(x,y)\simeq0.28$, slower and linear dynamics takes place in the tails of the excitation volume (turquoise points); intermediate regimes with progressively slower evolution towards saturation can also be observed in the slopes of the Rydberg profile (green and red points). We interpret the development of local saturation (significantly below $\rho_{r}^\mathrm{2D}(x,y)\simeq0.5$, expected for Rabi oscillating dynamics in non-interacting systems) as a direct consequence of the Rydberg blockade effect~\cite{Robicheaux2005,Stanojevic2009}. Once a Rydberg state is excited, the long-range repulsive van der Waals interactions at play amongst $\ket{nS}$ Rydberg atoms affect all surrounding atoms by shifting their energy levels, effectively suppressing the probability to excite another such state within a critical distance, called blockade radius. Thus, we can infer from the measured Rydberg fraction that an average number of $3.6$ atoms is blockaded by each Rydberg excitation. In contrast to the local dynamics, the measured global number of Rydberg atoms $\langle N_{r} \rangle$ does not show saturation at long excitation times $t_{exc}$ (inset of Fig.~\ref{fig:interaction_effects}(b)). This can be attributed to the ongoing Rydberg excitation of a significant number of atoms in the tails of the distribution, where the evolution is slower and linear as full blockade has not been reached.

In order to reproduce the experimentally observed excitation dynamics and verify whether it can be described as a non-interacting collection of superatoms~\cite{Dicke1954}, we implement a Monte Carlo master equation simulation that accounts for the influence of the Rydberg blockade effect, due to which all atoms within a blockade radius follow common many-body dynamics driven by the collectively enhanced superatom Rabi frequency. The model starts from a randomized distribution of ground state atoms enclosed in a volume defined by the experimental geometry. Each atom presents an excitation probability $\rho_i$ that is determined by solving the optical Bloch equations with a dephasing $\gamma$ and an enhanced Rabi frequency $\sqrt{N_{\text{bl}}}\Omega_{\text{eff}}$, where $N_{\text{bl}}$ is the number of blockaded atoms per superatom. The latter is obtained self-consistently from the Rydberg blockade radius $R_{\text{bl}}=(C_6/\hbar\Delta\nu)^{1/6}$ which depends on the van der Waals interaction coefficient $C_6$~\cite{WalkerSaffman2008} and on the spectral width of the excitation $\Delta\nu$, given by the combination of the Fourier limited width, the laser linewidth and the power broadening of the enhanced Rabi frequency. Hence, the distribution of Rydberg atoms is sampled by exciting each atom with its probability $\rho_i$, unless they lay within the blockaded volume of an already excited atom where their excitation is suppressed, and averaging over 50 simulated samples.
We apply our model to describe the experimentally observed local Rydberg fraction, accounting in each bin for the geometry of the ground state atoms along the imaging direction and fitting the local effective Rabi frequency $\Omega_{\text{eff}}$, whereas the dephasing $\gamma/2\pi = \SI{0.25}{\mega\hertz}$ is considered as a global fixed parameter for the whole distribution. Good agreement to the dynamics measured by depletion imaging is found in all excitation regimes, as depicted with solid lines in Fig.~\ref{fig:interaction_effects}(b): in the tails of the cloud, the simulation reproduces the incoherent linear dynamics, as the dephasing notably exceeds the effective excitation driving rate $\gamma \gg \Omega_{\text{eff}}$; in the center, it captures the development of saturation of the Rydberg fraction towards $\rho_{r}^\mathrm{2D}(x,y)\simeq0.28$ due to the increasing Rydberg blockade effect.

Leveraging the results of the local fits of the excitation dynamics, the effective Rabi frequency $\Omega_{\text{eff}}(x,y)$ can be spatially reconstructed without assumptions on the blue laser beam profile at the atom position, as shown in Fig.~\ref{fig:interaction_effects}(c). The profile presents deviations from a Gaussian distribution due to aberrations arising during the propagation through the optical system that conveys and focuses the blue beam (initially Gaussian shaped) onto the atom cloud.  Furthermore, this information on $\Omega_{\text{eff}}(x,y)$ can be employed as an input parameter of the Monte Carlo simulation to infer a realistic three-dimensional distribution of the Rydberg atom cloud for a given spatial arrangement of the ground state atoms, as illustrated in Fig.~\ref{fig:interaction_effects}(d). In each individual configuration all Rydberg atoms are separated by the blockade radius at high densities, thus accounting for the emergent spatial order in Rydberg gases~\cite{Reetz-Lamour2008,Loew2009,Dudin2012a,Schauss2012}. This distribution can be employed to model the spin dynamics of a spin 1/2 system without any additional free parameters~\cite{Schachenmayer2015a,Signoles2019,Hazzard2014a}. 

\section{Conclusion}

In this work we have demonstrated the use of depletion imaging to characterize the spatial properties of ultracold Rydberg gases. As this technique is based on well-established absorption imaging, it benefits from its strengths, offering the possibility to precisely calibrate the local number of Rydberg excitations, which is a notoriously difficult task by means of, e.g., ionization-based techniques due to the difficulty of determining absolute detection efficiencies. Depletion imaging can be performed within a narrow time interval of only a few microseconds, short compared to the usual lifetime of Rydberg states, thus leaving the Rydberg population unperturbed. Therefore, this measurement tool can be used in combination with other techniques such as state-selective field ionization~\cite{Guertler2004}, granting complementary knowledge. 

Depletion imaging provides local information about the two-dimensional Rydberg density distribution. Applied to the presented experiments, we found a saturation as a consequence of collective excitation dynamics. This technique may become an excellent diagnostic and probing tool for future local studies of Rydberg excitation scaling dynamics~\cite{Heidemann2007,Loew2007}, allowing to investigate a large parameter space where a rich phase diagram is expected~\cite{Valado2013,Helmrich2016}. Assisted by a Monte Carlo superatom model, we derived the discrete three dimensional distribution of the system. This can serve as the basis to study the properties of quantum spin systems, such as the phase diagram of the Heisenberg model~\cite{Dmitriev2002}. Extending the technique by a three-dimensional reconstruction method based on the underlying symmetry of the system can reveal the actual three-dimensional Rydberg distribution, allowing to study spatially resolved many-body dynamics of Ising spin systems in higher dimensions~\cite{Loew2009,Helmrich2018}.


\section*{Acknowledgments}
This work is part of and supported by the DFG Collaborative Research Centre "SFB 1225 (ISOQUANT)", the DFG Priority Program "GiRyd 1929" and WE2661/10.2, the European Union H2020 projects FET Proactive project RySQ (Grant No. 640378) and FET flagship project PASQuanS (Grant No. 817482) and the Heidelberg Center for Quantum Dynamics. V.G. acknowledges support from the IMPRS-QD. T.F. acknowledges funding by a graduate scholarship of the Heidelberg University (LGFG) and  R.F.A. from the Brazilian fund Ci\^{e}ncia sem Fronteiras.

\section*{References}
\bibliographystyle{iopart-num}
\bibliography{biblio} 

\end{document}